**Comments on the BCBS proposal for a New Standardized Approach for Operational Risk**

Giulio Mignola[1,*], Roberto Ugoccioni[1,*], Eric Cope[2,*]

(June 2016)

Abstract

On March 4th 2016 the Basel Committee on Banking Supervision published a consultative document where a new methodology, called the Standardized Measurement Approach (SMA), is introduced for computing Operational Risk regulatory capital for banks. In this note, the behavior of the SMA is studied under a variety of hypothetical and realistic conditions, showing that the simplicity of the new approach is very costly on other aspects: we find that the SMA does not respond appropriately to changes in the risk profile of a bank, nor is it capable of differentiating among the range of possible risk profiles across banks; that SMA capital results generally appear to be more variable across banks than the previous AMA option of fitting the loss data; that the SMA can result in banks over- or under-insuring against operational risks relative to previous AMA standards. Finally, we argue that the SMA is not only retrograde in terms of its capability to measure risk, but perhaps more importantly, it fails to create any link between management actions and capital requirement.

---

[1] Intesa Sanpaolo, Enterprise Risk Management Department
[2] Credit Suisse, Operational Risk Management – Capital
[*] The views and statements expressed in this article are those of the authors and do not necessarily reflect the views of Intesa Sanpolo Spa and its affiliates ("Intesa Sanpaolo") or Credit Suisse Group AG and its affiliates ("Credit Suisse"). Intesa Sanpaolo and Credit Suisse provide no guarantee with respect to the content and completeness of the article and disclaims responsibility for any use of the article.

Introduction

On March 4$^{th}$ 2016 the Basel Committee on Banking Supervision published a consultative document aimed to revise the entire minimum regulatory capital framework for operational risk (BCBS, 2016). This document proposes to replace all the currently available options for computing regulatory capital (BIA, TSA-ASA, AMA) with a single standardized measure called the Standardized Measurement Approach (SMA). The goal of the SMA is to enhance the simplicity of the capital calculation and promote greater comparability among banks, while retaining a degree of risk-sensitivity.

The SMA is undeniably simple. However, we take issue with the notion that it promotes comparability across banks, nor is the SMA as risk-sensitive as we might desire it to be. We base this evaluation on a study of the behavior of the SMA under a variety of hypothetical and realistic conditions, from which we derive several conclusions.

First, the SMA does not respond appropriately to changes in the risk profile of a bank, nor is it capable of strongly differentiating among the range of possible risk profiles across banks. For one, the SMA does not grow in proportion to expected operational losses, as one would expect of a capital measure. At most, there can be a 50% difference in capital requirements between two banks of similar size and yet at the extreme ends of the possible loss profiles under the SMA. Under the AMA or similar Value-at-Risk (VaR) type models, there would be more than a factor of 30 difference between the capital levels at these banks, indicating that the SMA is not appropriately risk-sensitive.

Second, SMA capital results generally appear to be more variable across banks than a simple AMA-type risk model would be that has been fitted to the loss data. The SMA is closely associated with a measure of bank income, which is imperfectly correlated with operational losses. Based on industry survey data collected from the ORX consortium (ORX, 2016), we find that the range of income levels observed together with a given operational loss profile can vary quite widely. This high degree of variability indicates that the SMA cannot achieve comparability, as banks with similar risk profiles may be assigned quite different levels of capital under the SMA.

Third, the SMA can result in banks over- or under-insuring against operational risks relative to AMA-level standards. Note that the SMA is not tied to any risk measure or standard, as the AMA measure was tied to the 99.9$^{th}$ percentile VaR of the annual total loss distribution. In our investigations, we find an extreme range of possible confidence levels associated with the SMA, which can be as low as the 99$^{th}$ percentile and as the high as the 99.9999$^{th}$ percentile, or more. This provides further evidence that the method does not sufficiently differentiate among risk profiles.

Description of the SMA approach

The main component of the SMA is a measure of business volume called the Business Indicator (BI), which is based on the main elements of Gross Income (today's driver for the BIA and the TSA), mingled with some indicators of expenses (including operational losses). The BI is composed of the sum of the 3-year averages of:

- **the Interest, Lease and Dividend Component (ILDC)**, which is basically the Net Interest margin, including net operating and financial lease results and dividend income;
- **the Service Component (SC),** which includes the maximum of Fee Income and Fee Expenses, as well as the maximum of Other Income and Other Expenses;
- **the Financial Component (FC),** which includes the absolute value of the P&L of the Trading Book and the absolute value of the P&L of the Banking Book.

That is, the BI can be expressed thus:

$$BI = ILDC_{avg} + SC_{avg} + FC_{avg}$$

The BI represents the single determining factor of the BI Component (BIC), which is the base value of the SMA. The BIC is obtained by applying a specific coefficient plus an offset to the BI. The coefficient and the offset depend on the size bucket of the BI as shown in the following table.

| BI range (€) | BIC Formula |
| --- | --- |
| 0 – 1 bn | 0.11·BI |
| 1 – 3 bn | 110m + 0.15·(BI - 1 bn) |
| 3 – 10 bn | 410m + 0.19·(BI - 3 bn) |
| 10 – 30 bn | 1.74bn + 0.23·(BI - 10 bn) |
| > 30 bn | 6.34bn + 0.29·(BI - 30 bn) |

**Table 1.** BI component of the SMA as a function of the BI indicator. The progressive behavior (more than linear) is explicit in the different coefficients applied to the different buckets of BI.

The piecewise linear relationship between the BI and the BIC is shown in the following graph, which shows the progressive (super-linear) nature of the BIC in relation to the BI.

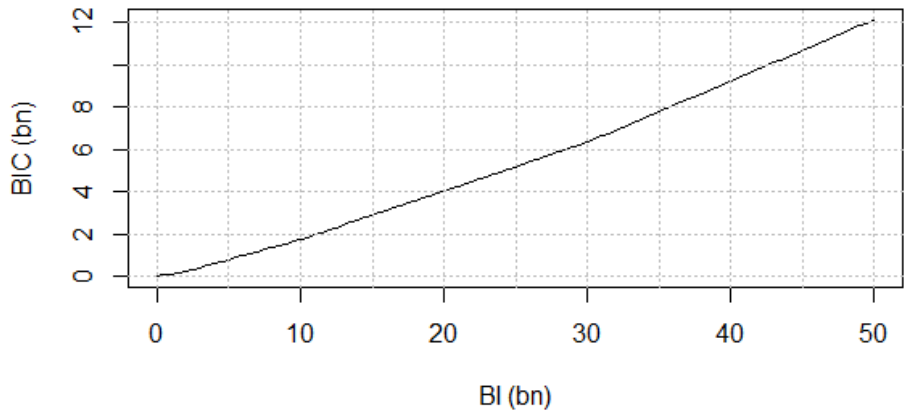

**Figure 1.** BI component of the SMA as a function of the BI indicator. The progressive behavior (more than linear) is clearly seen from the graph.

To supplement the BIC and take into account the different risk profiles of banks of similar size, the BCBS introduced a modifier to the BIC based on individual bank's internal losses termed the Loss Component (LC). The LC is required only for banks having a BI greater than €1 bn and is calculated according to the following expression:

$$LC = \frac{7 \cdot \sum_{10 \text{ yrs}} x + 7 \cdot \sum_{10 \text{ yrs}} x \cdot 1\{x > €10\text{m}\} + 5 \cdot \sum_{10 \text{ yrs}} x \cdot 1\{x > €100\text{m}\}}{10},$$

where $x$ represents individual losses experienced over a period of 10 years, and $1\{\cdot\}$ represents the indicator function, which equals 0 if the condition inside the brackets is false, and 1 otherwise.

The LC can be expressed as a multiple of the total annual expected loss, as follows:

$$LC = \alpha^* \cdot EL,$$

where EL is the expected loss and the risk factor $\alpha^*$ is given by

$$\alpha^* = 7 + 7\alpha_{10} + 5\alpha_{100}.$$

Here, $\alpha_{10}$ and $\alpha_{100}$ are the fractions of total losses exceeding 10 or 100 million, respectively, i.e.,

$$\alpha_Y = \frac{\sum x \cdot 1\{x > Y\}}{\sum x}$$

The risk factor $\alpha^*$ can take values as low as 7 – for banks that have only experienced losses smaller than €10mn (i.e., $\alpha_{10} = \alpha_{100} = 0$) – or as high as 19, for banks that have only experienced losses greater than €100mn (i.e., $\alpha_{10} = \alpha_{100} = 1$). These extreme outcomes define the range of loss profiles that can be differentiated by the LC; hence $\alpha^*$ represents the SMA's measure of the riskiness of the bank.

The LC enters into the SMA as a multiplier of the BIC for banks having a BI greater than €1 bn. (The LC does not play any role for banks with BI less than €1bn.) The exact formula is

$$SMA = \begin{cases} BIC & \text{if } BI < €1\text{bn} \\ €110\text{mn} + (BIC - €110\text{mn}) \cdot \ln\left(e - 1 + \frac{LC}{BIC}\right) & \text{if } BI > €1\text{bn} \end{cases}$$

For banks with BI greater than €1bn, if the LC is larger (resp. smaller) than the BIC, the SMA will be larger (smaller) than the BIC. When the LC is exactly equal to the BIC (as the BCBS claims will be true for the "industry average bank"), then the SMA will also coincide with the BIC.

Considerations on the SMA

The objective of the minimum capital requirement should be to protect the bank from the effects of unexpected losses, creating a buffer in term of capital that would be enough to absorb those losses with a reasonable degree of certainty. In the current AMA framework, this degree of certainty is defined as the 99.9% VaR level of confidence in the annual total loss distribution. As mentioned above, the SMA is not directly linked to any such risk measure or standard, and the operational loss profile is explicitly represented through the Loss Component, which as we have seen can be reduced to an expected loss component (EL) times a risk factor $\alpha^*$. The risk factor increases if a bank changes from having a predominantly high-frequency, low-intensity (HFLI) loss profile, to a more low-frequency, high-intensity (LFHI) loss profile.

Consistent with the objective stated above, we should expect the SMA (or any reasonable measure of capital under a heavy-tailed loss regime) to have the following properties:

1. A bank with high EL should have significantly more capital than a bank with low EL, assuming the risk factors at these banks are the same.
2. A bank with a LFHI risk factor should have significantly more capital than a bank with an HFLI risk factor, assuming that the EL at these banks is the same.

However, the SMA obeys neither of these properties. The main component of the SMA is clearly the size of the business rather than the loss profile. In fact, the role played by losses is rather mild overall: the effect of an 100% increase in the Loss Component would be no greater than a 33% increase in the SMA, assuming no change in BI. If this change in the Loss Component is strictly due to a change in the EL, and the BI and the risk factor $\alpha^*$ do not change, then the ratio of the SMA to the EL decreases substantially, an effect that appears to be in contradiction with the first principle listed above. More generally, Figure 2 indicates how the ratio of the SMA to the EL changes as the Loss Component changes with respect to the BI Component across the range of possible values of $\alpha^*$. In all cases, the decrease is very rapid, as the ratio SMA/EL decreases from 30 at LC/BIC =0.5 down to 12.5, at the point where LC/BIC=2 and $\alpha^*$ takes its worst-case value of 19. For a lower internal loss multiplier, e.g. $\alpha^*$ = 7, the SMA/EL ratio decreases from 11 to less than 5 in the same range of variation for LC/BIC.

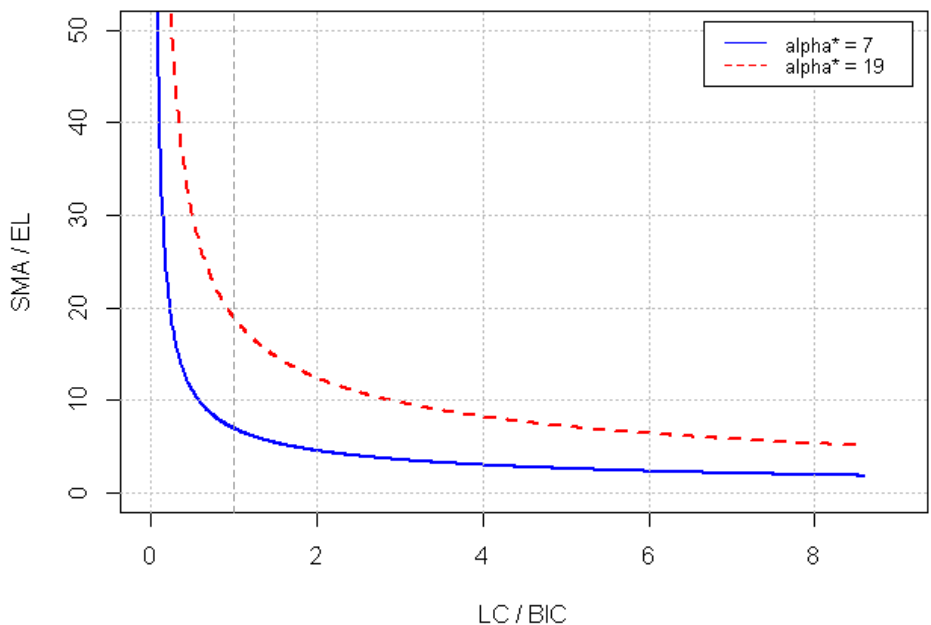

**Figure 2.** SMA/EL ratio for a given value of the multiplier $\alpha^*$ as a function of the ratio of the EL component to the BI component. In red (dashed line) the maximum possible value of 19 (i.e. all losses > 100m). In blue (solid line) the lowest possible value of 7 (i.e. all losses < 10 m). The vertical line at 1 identifies the "average" bank. This plot of course applies for BI > 1 bn.

Another interesting aspect is the effect of different risk factors $\alpha^*$. Consider two banks having the same BI and the same EL, but where one bank has the minimum risk factor of 7 and the other the maximum value of 19. The ratio between the SMA values of these two banks is bounded above by 1.5 and depends on the LC/BIC value. Figure 3 indicates that the maximum ratios between the SMA values of these banks is obtained when the ratio of the EL to the BIC is around 0.2, corresponding to an LC/BIC ratio of about 1.4 for banks with a low risk profile ($\alpha^* = 7$). After this point, the effect decreases as the EL increases. This is quite counterintuitive given the second property of reasonable capital measures listed above: banks with a high expected loss and a high proportion of large losses should be capitalized significantly more in relation to banks with a lower EL.

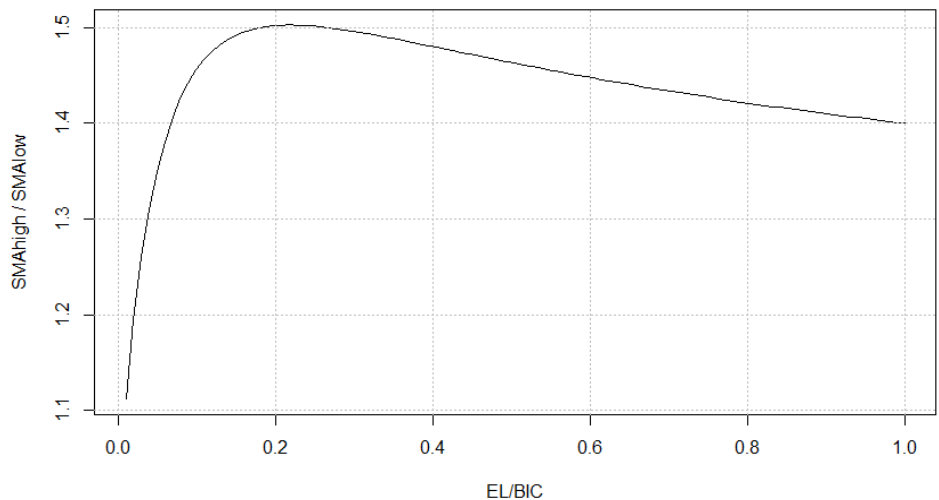

**Figure 3.** Ratio of the SMA value computed with the higher Internal Loss multiplier of 19 to the SMA calculated with the lower Internal Loss multiplier of 7 as a function of the EL/BIC. Reasonable range of EL/BIC is between 1% and 50%; larger values are not excluded but should be rare (these would be very risky banks indeed).

Theoretical case studies

In order to understand the level of variability – and hence the consistency – of the SMA, we have run a number of simulation experiments. The simulations are based on realistic assumptions regarding the frequency and severity of loss-generating processes commonly observed in the industry, as well as on the typical levels of the Business Indicator associated with banks having those loss profiles. These studies show that not only is the SMA highly variable in relation to a simple internal model-based approach to measuring capital, but also that most banks will likely be over-insured against operational risk in comparison to the prior AMA standard, sometimes greatly so.

Consider a very simple bank, with just one loss-generating mechanism (i.e., defined by one loss severity model and one loss frequency model) where all the parameters of these distributions are known, so that the associated capital requirement at the 99.9$^{th}$ percentile of the annual aggregated loss distribution can be easily derived. In this simple but sufficiently realistic case, the bank can be described using three parameters: its average frequency $\lambda$, assuming a Poisson distribution; and the shape $\sigma$ and location $\mu$ of its severity distribution, assuming a lognormal distribution. Using the standard method of the Loss Distribution Approach (LDA), it is possible to aggregate the frequency and severity distributions to arrive at the AMA capital requirement. In addition, the distribution of the Loss Component of the SMA can also be derived.

To fully compute the SMA capital requirement, however, we must additionally make certain assumptions regarding the size of the Business Indicator and its relationship to the Loss Component. Based on a survey of 54 banks conducted by the ORX consortium, we have data on the Loss Component and Business Indicator Components over a three-year period from 2013-2015. There is a strong correlation between the two components, as shown in the graph below.

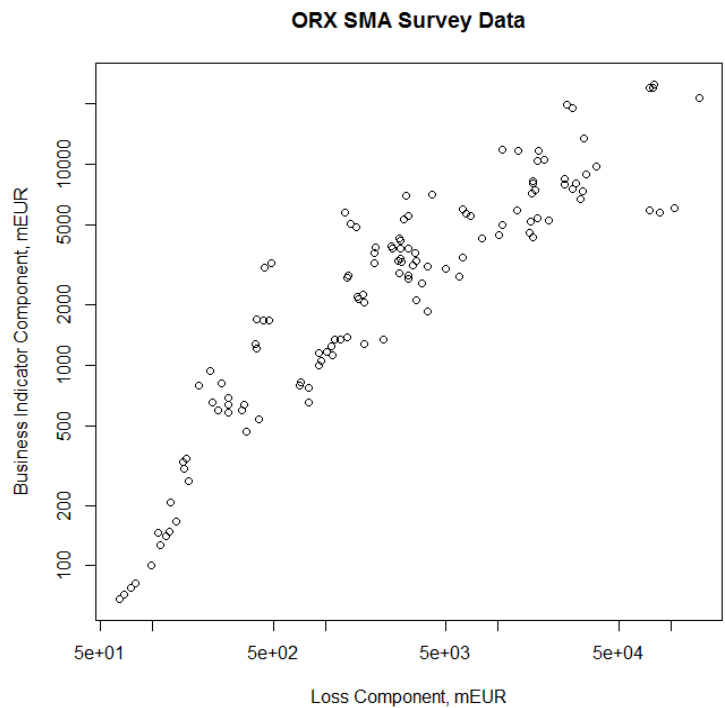

**Figure 4.** ORX survey data of BIC and LC values over a three-year period, collected from 54 consortium members.

Using a simple transformation of the data, it is possible to determine a linear relationship between these two components. A ordinary least-squares regression of the log of the BI Component against the log of the log of the Loss Component, as shown in the graph at left below, indicates that the residuals are fairly Normal in character, as evidenced by the quantile-quantile plot of the regression residuals, below right.

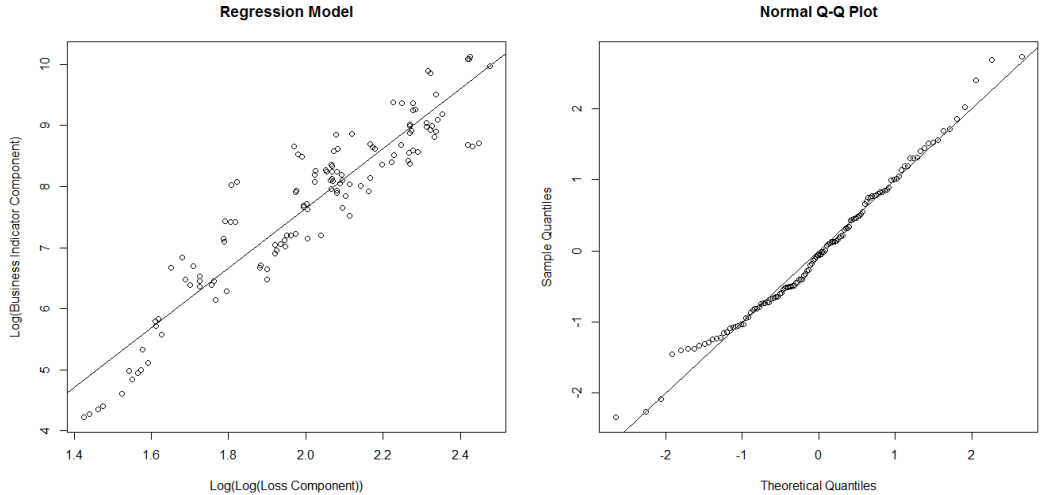

**Figure 5.** Fitted regression of log(BIC) on log(log(LC)) (left graph) and quantile-quantile plot of residuals against a Normal distribution (right graph).

Various statistical tests of these residuals (Anderson-Darling, Shapiro-Wilk, Cramer-von Mises, Lilliefors, etc.) all pass at the 5% significance level, confirming that the residuals are reasonably consistent with a Normal distribution. The fitted regression is expressed as

$$\log(BIC) = -2.16 + 4.90 \cdot \log(\log(LC)) + \varepsilon, \qquad (1)$$

where $\varepsilon$ is a Normal random variable with mean 0 and standard deviation 0.486. We have used this regression relation to simulate realistic values of the BI Component, given the value of the expected Loss Component derived from the assumed parameters.

In order to ensure that the assumed Poisson and Lognormal parameters are themselves within a reasonable range, we assume that the following relationships hold, in keeping with the ranges of these values reported by ORX.

A. The AMA capital level (99.9$^{th}$ percentile of the annual aggregate loss distribution) implied by the parameters exceeds €10mn and is less than €50bn.
B. The expected value of the Loss Component exceeds €64mn and is less than €150bn.
C. The ratio of the frequency of losses exceeding €20,000 to the median Business Indicator (determined by the BI Component from the regression line, expressed in billions of euro) exceeds 20.

The simulation procedure was executed as follows:

1. Select values of the Poisson parameter λ and the lognormal parameters μ and σ.
2. Determine the implied AMA capital level, expected Loss Component, median BI component, and frequency of losses exceeding €20,000.
3. If the three conditions (A)-(C) hold, then run the following procedure 100 times:
    a. Simulate 10 years of loss data based on the assumed Poisson and Lognormal distributions

b. Compute the Loss Component from this data
c. Generate a BI Component based the expected Loss Component using the regression equation (1)
d. Compute the SMA requirement based on the Loss Component and BI Component
e. Fit a Lognormal and Poisson distribution to the loss data exceeding €10,000 (the data collection floor for the SMA), and compute the AMA capital levels implied by the estimated parameters. The capital levels were determined using the single-loss approximation (SLA) formula

$$SLA = \hat{F}^{-1}\left(1 - \frac{1 - 0.999}{\hat{\lambda}}\right) + \hat{\lambda} \cdot \mu_{\hat{F}},$$

where $\hat{F}$ represents the fitted lognormal distribution (left-truncated at €10,000), $\mu_F$ is the mean of this distribution, and $\hat{\lambda}$ is the estimated frequency of losses exceeding €10,000.

4. Based on the 100 realizations of the SMA and internal-model based capital (which we shall refer to as SLA capital), compute the following quantities of interest:
   a. The coefficient of variation of the SMA capital, i.e., the standard deviation divided by the mean of the 100 realizations.
   b. The coefficient of variation of the SLA capital.

In the simulations, we allowed the value of μ to range between 8 and 12, σ to range between 1 and 4, and λ to be in the set {100, 500, 1000, 5000, 10000}. (Note that λ represents the mean frequency of all losses, no matter how small. The number of losses used in fitting the parameters in step (3e) of the simulation procedure was based only on the realized losses exceeding €10,000, which was usually a much lower number than λ, depending on the value of μ and σ. In all, 315 combinations of values were tested, of which only 134 met conditions (A)-(C).

The results indicate that the SMA results are generally considerably more variable than the internal model based SLA results. The graph below shows that the coefficient of variation of the SMA is consistently well above that of the SLA, for all values of the (expected) Loss Component in the range observed in the industry. The primary reason for the high variability of the SMA is that the BI Component is so variable – i.e., banks of many sizes can have a similar Loss Component – that it cannot serve as a highly reliable indicator of risk. By contrast, the simple internal model-based outcomes show better overall variability.

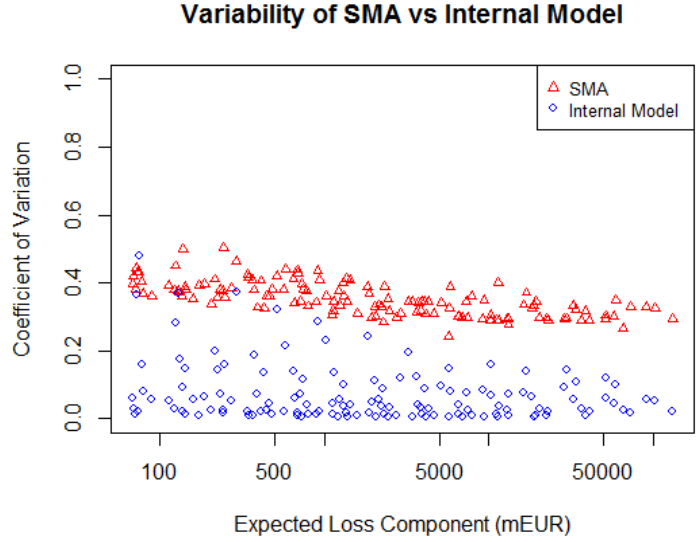

**Figure 6.** Simulation results comparing the coefficient of variation of the SMA results for each parameter setting to the results based on the internal model fit.

We can obtain a more granular view of the relation between the SMA and the true VaR capital at the 99.9$^{th}$ percentile by examining a range of numeric results for certain specific combinations of parameters. A second set of results were determined according to a slightly modified procedure. First, a certain level of the Business Indicator (in billions) is assumed, using the example of a small bank (BI of €8bn), a medium bank (BI of €20bn), and a large bank (BI of €40bn); these selections also determine the BIC. Next, the expected value of the Loss Component is determined using equation (1), considering three different quantiles of the stochastic (Normal) component (10%, 50% and 90%), thus obtaining a low, a median and a high LC/BIC ratio. For each of these 9 combinations of BI and LC, the parameters μ and σ are chosen in order to assess three different regimes of the $\alpha^*$ parameter: a low value (around 8), a medium value (around 11) and a high value (around 15) so that situations dominated by losses below €10m up to those with large losses are considered. For a given value of $\alpha^*$, two combinations of μ and σ are also selected in order to have a wide range of possible frequencies. Based on the LC and $\alpha^*$ parameter, we may of course determine the values of EL and λ.

To understand level of insurance that the SMA is providing against operational losses in each case, denote by $F_A(SMA)$ the value of the cumulative distribution function (cdf) of the aggregate annual loss distribution at the level of the SMA. We find it convenient to represent $F_A(SMA)$ in terms of the "number of 9's" in the decimal representation. For example, the standard AMA level measures the 0.999 confidence level of annual total losses, which is the "three-9's" level. If the SMA corresponds to the 0.99 or 0.9999 level, this would be "two-9's" and "four-9's" respectively. In general, the "number of 9's" measure is determined by the following formula:

$$\#9's = -\log_{10}\bigl(1 - F_A(SMA)\bigr).$$

The specific values of the input parameters that were tested (BI, $\mu$, $\sigma$ and the quantile) are displayed in the left-hand columns in Table 2 below. From these values, all other columns are computed; note that all of these quantities are derived conditional on the losses exceeding 10,000 euro, in keeping with the SMA data requirements. (In the table, rows which would be rejected under condition (C) are printed in have gray text using italics.) In this case, the 99.9$^{th}$ percentile (referred to as VaR in the table) was determined by using the Fast Fourier Transform (FFT). The final columns indicates the ratio of the SMA to the true 99.9$^{th}$ percentile (SMA / VaR) and the number of 9's associated with the SMA. These quantities are direct indications of the capability of the SMA to capture the theoretical risk profile.

Reviewing these results, it is clear that the SMA is materially overestimating the theoretical VaR for small values of $\alpha^*$ (values close to the theoretical minimum of 7). Conversely, when $\alpha^*$ is large, in the range of 14-16, the SMA is generally underestimating VaR. Many of these cases however did not meet condition (C), and therefore may not correspond to actual banks. Among the median values (10-12) of $\alpha^*$, we do not observe perfectly consistent results, although in the majority of cases, the SMA overestimates VaR.

The "number of 9's" associated with the SMA percentile level in this experiment ranges widely, with some outcomes below the two-9's level, indicating that the SMA would represent only about the once-per-100-year event, and other outcomes above the six-9's level, in which case the bank is quite heavily over-insured against operational risk.

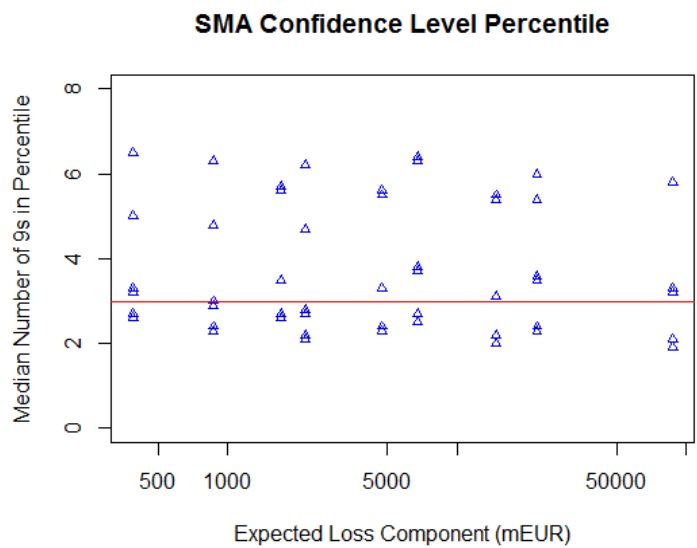

**Figure 7.** Simulation results showing the median "number of 9's" in the decimal representation of the SMA percentiles with respect to the true annual aggregate loss distribution, for the test cases in Table 2. The red line indicates the three-9's level at which the AMA is calibrated.

In addition, the primary driver of the SMA is clearly bank size (income), whereas the VaR measure is mainly driven by the loss profile. As a result, there is a range of risk factors $\alpha^*$ that are consistent with a fixed set of BI and LC values, indicating that a wide variety of risk profiles, with very different levels of VaR, are consistent with the same level of the SMA.

**Table 2.** Comparison between the SMA and the theoretical capital requirement (VaR) under a range of situations.

| BI & BIC (bn) | μ | σ | quantile | LC(bn) | λ | α* | EL(bn) | SMA(bn) | VaR(bn) | SMA/VaR | #9's |
|---|---|---|---|---|---|---|---|---|---|---|---|
| 8<br>(1.36) | 8 | 2 | low | 0.390 | 735 | 7.2 | 0.055 | 0.980 | 0.116 | 8.5 | 6.5 |
| | | | median | 0.876 | 1,650 | | 0.122 | 1.185 | 0.207 | 5.7 | 6.3 |
| | | | high | 2.195 | 4,133 | | 0.307 | 1.615 | 0.427 | 3.8 | 6.2 |
| | 9 | 2.1 | low | 0.390 | 323 | 7.7 | 0.050 | 0.980 | 0.203 | 4.8 | 5.0 |
| | | | median | 0.876 | 724 | | 0.113 | 1.185 | 0.329 | 3.6 | 4.8 |
| | | | high | 2.195 | 1,814 | | 0.283 | 1.615 | 0.601 | 2.7 | 4.7 |
| | 9 | 2.6 | low | 0.390 | 71 | 10.9 | 0.036 | 0.980 | 0.709 | 1.4 | 3.3 |
| | | | median | 0.876 | 159 | | 0.080 | 1.185 | 1.140 | 1.0 | 3.0 |
| | | | high | 2.195 | 398 | | 0.201 | 1.615 | 1.942 | 0.8 | 2.8 |
| | 10 | 2.5 | low | 0.390 | 42 | 11.6 | 0.034 | 0.980 | 0.786 | 1.2 | 3.2 |
| | | | median | 0.876 | 94 | | 0.075 | 1.185 | 1.259 | 0.9 | 2.9 |
| | | | high | 2.195 | 236 | | 0.189 | 1.615 | 2.128 | 0.8 | 2.7 |
| | 9 | 3.1 | low | 0.390 | *12* | 15.2 | *0.026* | *0.980* | *1.729* | *0.6* | *2.7* |
| | | | median | 0.876 | *28* | | *0.058* | *1.185* | *3.123* | *0.4* | *2.4* |
| | | | high | 2.195 | 69 | | 0.145 | 1.615 | 5.950 | 0.3 | 2.2 |
| | 10 | 3.1 | low | 0.390 | *5* | 16.3 | *0.024* | *0.980* | *2.057* | *0.5* | *2.6* |
| | | | median | 0.876 | *12* | | *0.054* | *1.185* | *3.846* | *0.3* | *2.3* |
| | | | high | 2.195 | *30* | | *0.134* | *1.615* | *7.579* | *0.2* | *2.1* |
| 20<br>(4.04) | 9 | 2.1 | low | 1.722 | 1,424 | 7.7 | 0.222 | 3.108 | 0.509 | 6.1 | 5.7 |
| | | | median | 4.728 | 3,909 | | 0.610 | 4.279 | 1.044 | 4.1 | 5.6 |
| | | | high | 14.887 | 12,305 | | 1.921 | 6.740 | 2.606 | 2.6 | 5.5 |
| | 10 | 2 | low | 1.722 | 862 | 8.1 | 0.213 | 3.108 | 0.552 | 5.6 | 5.6 |
| | | | median | 4.728 | 2,367 | | 0.584 | 4.279 | 1.094 | 3.9 | 5.5 |
| | | | high | 14.887 | 7,450 | | 1.840 | 6.740 | 2.639 | 2.6 | 5.4 |
| | 9 | 2.6 | low | 1.722 | 312 | 10.9 | 0.158 | 3.108 | 1.688 | 1.8 | 3.5 |
| | | | median | 4.728 | 857 | | 0.434 | 4.279 | 3.033 | 1.4 | 3.3 |
| | | | high | 14.887 | 2,699 | | 1.365 | 6.740 | 6.002 | 1.1 | 3.1 |
| | 10 | 2.5 | low | 1.722 | 185 | 11.6 | 0.148 | 3.108 | 1.853 | 1.7 | 3.5 |
| | | | median | 4.728 | 508 | | 0.407 | 4.279 | 3.295 | 1.3 | 3.3 |
| | | | high | 14.887 | 1,598 | | 1.281 | 6.740 | 6.407 | 1.1 | 3.1 |
| | 9 | 3.1 | low | 1.722 | *54* | 15.2 | *0.114* | *3.108* | *5.031* | *0.6* | *2.7* |
| | | | median | 4.728 | 149 | | 0.312 | 4.279 | 10.012 | 0.4 | 2.4 |
| | | | high | 14.887 | 470 | | 0.982 | 6.740 | 21.240 | 0.3 | 2.2 |
| | 10 | 3.1 | low | 1.722 | *24* | 16.3 | *0.105* | *3.108* | *6.356* | *0.5* | *2.6* |
| | | | median | 4.728 | *65* | | *0.289* | *4.279* | *13.062* | *0.3* | *2.3* |
| | | | high | 14.887 | 203 | | 0.911 | 6.740 | 28.533 | 0.2 | 2.0 |
| 40<br>(9.24) | 9 | 2.1 | low | 6.781 | 5,605 | 7.7 | 0.875 | 8.299 | 1.377 | 6.0 | 6.4 |
| | | | median | 22.418 | 18,531 | | 2.893 | 13.091 | 3.697 | 3.5 | 5.4 |
| | | | high | 87.153 | 72,041 | | 11.246 | 22.127 | 12.497 | 1.8 | 5.8 |
| | 10 | 2 | low | 6.781 | 3,394 | 8.1 | 0.838 | 8.299 | 1.426 | 5.8 | 6.3 |
| | | | median | 22.418 | 11,220 | | 2.771 | 13.091 | 3.706 | 3.5 | 6.0 |
| | | | high | 87.153 | 43,618 | | 10.773 | 22.127 | 12.206 | 1.8 | 5.8 |
| | 9 | 2.6 | low | 6.781 | 1,229 | 10.9 | 0.622 | 8.299 | 3.748 | 2.2 | 3.8 |
| | | | median | 22.418 | 4,064 | | 2.056 | 13.091 | 7.726 | 1.7 | 3.6 |
| | | | high | 87.153 | 15,801 | | 7.992 | 22.127 | 18.838 | 1.2 | 3.3 |
| | 10 | 2.5 | low | 6.781 | 728 | 11.6 | 0.583 | 8.299 | 4.052 | 2.0 | 3.7 |
| | | | median | 22.418 | 2,407 | | 1.929 | 13.091 | 8.185 | 1.6 | 3.5 |
| | | | high | 87.153 | 9,356 | | 7.499 | 22.127 | 19.377 | 1.1 | 3.2 |
| | 9 | 3.1 | low | 6.781 | *214* | 15.2 | *0.447* | *8.299* | *12.720* | *0.7* | *2.7* |
| | | | median | 22.418 | 708 | | 1.479 | 13.091 | 27.615 | 0.5 | 2.4 |
| | | | high | 87.153 | 2,752 | | 5.750 | 22.127 | 64.962 | 0.3 | 2.1 |
| | 10 | 3.1 | low | 6.781 | *93* | 16.3 | *0.415* | *8.299* | *16.763* | *0.5* | *2.5* |
| | | | median | 22.418 | 306 | | 1.372 | 13.091 | 37.415 | 0.3 | 2.3 |
| | | | high | 87.153 | 1,191 | | 5.334 | 22.127 | 89.801 | 0.2 | 1.9 |

Finally, we can observe a strong relationship between VaR and EL, given the value of the risk factor $\alpha^*$, in the examples shown in Table 2. Figure 8 plots these outcomes on a log-scale, with regression lines superimposed corresponding to each value of $\alpha^*$, estimated so as to have a common slope. The distance between the upper and lower lines is 3.5, which indicates that the VaR level when $\alpha^*$ = 16.3 is about exp(3.5) = 33 times the value of the VaR level when $\alpha^*$ = 7.2. Compare this value with the factor of 1.5 indicated earlier as the maximum possible difference between the worst-case and best-case risk profiles under the SMA.

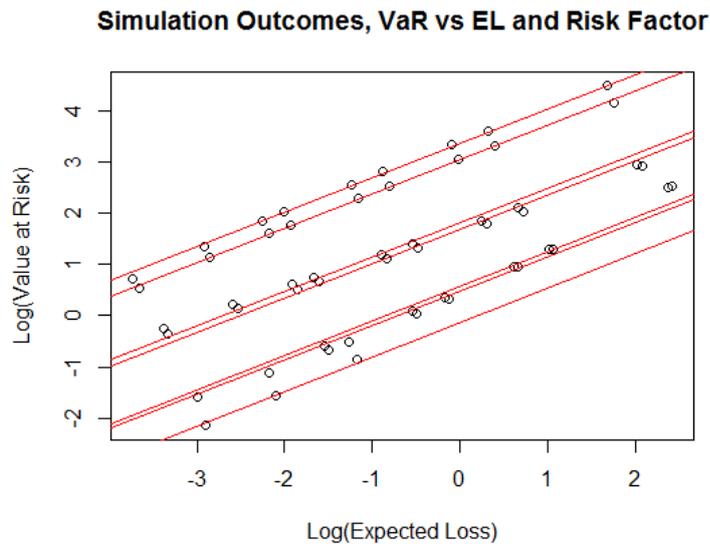

**Figure 8.** Plot of VaR against EL values from Table 2, plotted on a log scale. Each line represents a regression of log(VaR) against log(EL) for a different value of $\alpha^*$, which ranges in the set {7.2, 7.7, 8.1, 10.9, 11.6, 15.2, 16.3} in the examples. The lowest regression line corresponds to $\alpha^*$ = 7.2, and the highest to $\alpha^*$ = 16.3.

Based on the fact that banks with a high value of $\alpha^*$, and hence a very high percentage of very large losses, have a substantially lower SMA compared to the theoretical risk profile expressed by the VaR, a systematic bias is introduced by the use of the SMA in the population of banks. The effect of this bias can be typically seen for large international banks that are exposed to large fines and penalties related to conduct risk[3]. These banks are mainly based in the US, or have a large US footprint. Thus, one would expect that the impact of the SMA on US banks would be minimal, or possibly even a reduction in capital, whereas the impact on other jurisdictions would be to increase the capital requirement, in some cases to a large degree. This expectation has recently been confirmed in the results of the cited ORX Survey (ORX, 2016).

---

[3] Fines of tens of billions of dollars have been common in the past years for cases related to subprime mortgages and violation of international sanctions or other financial misconduct.

Conclusions

While the SMA may represent an improvement over other, standardized methods for calculating capital, it still belongs squarely in the category of size-driven capital standards, as the influence of the loss profile is quite minimal. As the simulation tests have indicated, it is certainly not a highly risk-sensitive measure.

Replacing the AMA with the SMA would represent a material step backward in the capability of banks to effectively hold capital against their operational risk exposure. Although it was well-known (and often publicly expressed, see e.g. Cope et al., 2010) by most banks that the AMA suffers from serious problems, the ability of an internal models-based approach to setting capital is still far higher than that of the proposed SMA.

The SMA is not only retrograde in terms of its capability to measure risk, but perhaps more importantly, it fails to create any link between management actions and capital requirement. Consider again two banks of the same size and loss profile that discover a similar operational vulnerability. One of these banks decides to invest heavily in improvements (controls, resources, exiting legally risky environments etc.), while the second bank decides to do absolutely nothing. Under the SMA, both banks will continue to carry a similar capital requirement for many years. On the other hand, with the AMA, the first bank has the possibility to demonstrate the reduced risk profile through forward-looking internal estimates. While still not perfect, the AMA is at least directionally correct on this issue, while the SMA will reward bad practices and will penalize, in relative terms, good banks.

In most jurisdictions, the Pillar I minimum capital is considered as a floor for any Pillar II estimates. If the calibration of the SMA is kept at the proposed level, the use of internal models for Pillar II will also be hampered for low risk banks. Banks will therefore have very little incentive to invest in a costly Pillar II process if it will have no influence on the final binding level for the capital ratios.

Very few people would say that the AMA is a perfect method for determining capital. At bottom, it suffers from problems of design and implementation, which have led to the symptoms that are now leading the Basel committee towards adopting a universal standardized approach. In particular, the choice of $99.9^{th}$ percentile as the measurement standard essentially guaranteed that the bank-internal models would not return results that were either stable or comparable. Based on the amount of data that banks have, it is not feasible to estimate the once-per-thousand-year annual loss with any degree of precision. Moreover, given the difficulty of achieving stable or usable results, the models became more and more complex as banks continued to apply more conditions and selection criteria. In addition, regulators in different jurisdictions applied the Basel rules differently, creating global imbalances in capital levels.

Therefore it is no wonder that we observe today a lack of simplicity and comparability in the AMA models – this outcome was virtually guaranteed from the start. However, this should not be interpreted as an indictment of internal models generally. There have been many proposals put forward that could improve the accuracy and robustness of an internal models framework, including basing the capital on a more attainable measurement standard, developing industry benchmarks, restricting the range of modeling practice, and allowing causal or structural models to supplement the statistical VaR models. Just because the $99.9^{th}$ percentile cannot be reliably estimated from a limited supply of data does not mean that internal models should have no role to play in determining the capital charge.

It would be a mistake to revert to a fully standardized approach such as the SMA. As we have shown, this measure fails according to two of the three desired outcomes that the Basel Committee put forward – comparability and risk-sensitivity. The SMA only succeeds in being simple. One may however do well to question why simplicity is even an objective at all, given the breadth and variety of operational risks affecting banks today.